\documentclass[aps,prl,twocolumn,superscriptaddress,showpacs,amsmath,amssymb]{revtex4}

\usepackage[utf8]{inputenc}
\usepackage{graphicx}

\newcommand{\ud}{\mathrm{d}}
\newcommand{\ang}[1]{\langle #1 \rangle}

\newcommand{\ue}{\mathrm{e}}
\begin{document}

\title{Exact Moment Scaling from Multiplicative Noise}

\author{Giacomo Bormetti}
\email[Electronic address: ]{giacomo.bormetti@pv.infn.it}
\affiliation{CeRS - IUSS, V.le Lungo Ticino Sforza 56, Pavia, 27100, Italy}
\affiliation{INFN - Sezione di Pavia, via Bassi 6, Pavia, 27100, Italy}

\author{Danilo Delpini}
\email[Electronic address: ]{danilo.delpini@pv.infn.it}
\affiliation{Dipartimento di Fisica Nucleare e Teorica, Università degli Studi di Pavia}
\affiliation{INFN - Sezione di Pavia, via Bassi 6, Pavia, 27100, Italy}

\date{\today}

\begin{abstract}
  For a general class of diffusion processes with multiplicative noise, describing a variety of physical as well as financial 
  phenomena, mostly typical of complex systems, we obtain the analytical solution for the moments at all times.
  We allow for a non trivial time dependence of the microscopic  dynamics and we analytically characterize the process evolution, possibly towards a stationary state,
  and the direct relationship existing between the drift and diffusion coefficients and the time scaling of the moments.
\end{abstract}
\maketitle
Many different physical phenomena exhibit a complex behaviour characterized by long range correlations,
long time memory, scale invariance and the emergence of non Gaussian distributions associated
to their statistical description. Deviations from the Maxwell-Boltzmann statistics were usually
considered as a clear mark of an out of equilibrium system, but recently it has been recognized
that Normality is not the most general paradigm describing the equilibrium state. Indeed, in terms
of a microscopic description provided by the Langevin equation, power law tails stem naturally assuming
the damping coefficient to have a stochastic nature~\cite{Biro_Jakovac:2005}. From a macroscopic
point of view, the superposition of an additive Gaussian noise with a multiplicative one leads to
a Fokker-Planck (FP) equation with linear drift and quadratic diffusion coefficients.
Processes leading to a macroscopic equation with the same structure emerge in the description of several
physical systems ranging from turbulent velocity flows~\cite{Friedrich_Peinke:1997}, power law spectra in $e^+e^-$, $p\bar p$ and heavy ions 
collisions~\cite{Wilk:2000}, anomalous diffusion phenomena~\cite{Borland:1998}, to the study of non stationary scaling Markov processes with
Hurst exponent $H\neq 1/2$~\cite{McCauley_etal:2007}.
Moreover the same dynamics has been shown to describe heartbeat interval fluctuations, foreign exchange markets~\cite{Ghashemi_etal:2006},
option markets~\cite{BorlandPRL:2002} and the statistical features of medium-term log-returns
in a market with both fundamental and technical traders \cite{Shaw:2009}.
The explicit analytical characterization of the probability density function (PDF) for these processes has been carried out
only for the steady state~\cite{Biro_Jakovac:2005}, while an analytical description at finite time can be given only in
terms of a formal expansion on the set of eigenfunctions of the FP operator~\cite{Schenzle_Brand:1979}.
In this Report, we provide a complete description of these processes in terms of their moments, allowing for a more general
time dependence for both the drift and diffusion coefficients. We obtain
closed-form expressions for the moments at finite time; in particular, we are able to characterize analytically how they thermalize to
the stationary state and we highlight the existence of a direct, simple relationship between the function regulating the time dependence
and the scaling of the process over time.

\begin{figure}[t]
  \begin{center}
    \includegraphics[width=.41\textwidth]{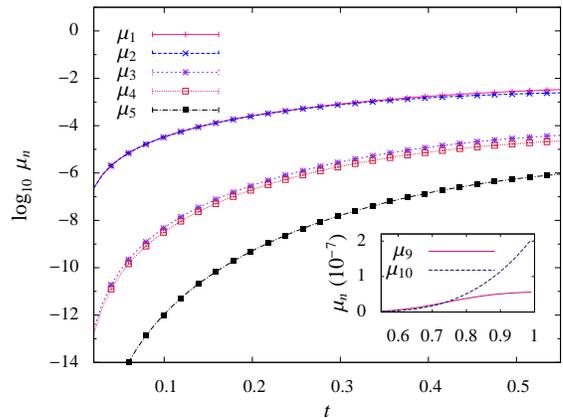}
  \end{center}
  \caption{(Color online) Lowest order moments for $a=-20$, $b=d=e=0.1$, $c=4.5$,
  with $g=t^{\beta}$, $\beta=2$ and $\tilde{p}_0(x)=\delta(x)$.
  In the inset the last converging moment is compared to the first diverging one.}
  \label{fig:PL10_stretched}
\end{figure}

\begin{figure}[t]
  \begin{center}
    \includegraphics[width=.41\textwidth]{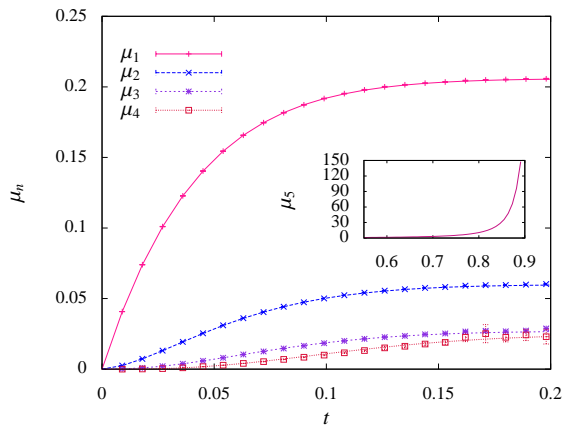}
  \end{center}
  \caption{(Color online) Lowest order moments for $a=-24.3$, $b=5$, $c=12.2$, $d=e=0.1$, with $t_\mathrm{lim}=1$ and $\tilde{p}_0(x)=\delta(x)$.
  The first diverging moment $\mu_5$ is shown in the inset.}
  \label{fig:PL5}
\end{figure}

We start from the stochastic differential equation (SDE) describing the microscopic dynamics under It\^o prescription~\cite{Risken:1996}
\begin{equation}
  \ud X_t = \frac{a X_t+b}{g(t)}\ud t + \sqrt{\frac{c X_t^2 + d X_t + e(t)}{g(t)}}~\ud W_t\ ,
  \label{eq:model}
\end{equation}
with initial time condition $X_{t_0}=X_{0}$,  $t_0\in D\subseteq [0,t_\mathrm{lim})$ with $t_\mathrm{lim}$ possibly $+\infty$;
$W_t$ is the standard Brownian motion, $a$, $b$, $c$, $d$ are real constants, $1/g(t)$ and $e(t)$ are non negative smooth functions of the time over $D$. 
For the diffusion coefficient to be meaningful, it has to satisfy $d^2-4ce(t) \leq 0$ with $c\geq 0$.
Application of the It\^o Lemma to $f(X_t)=X_t^n$ leads to the following integral relation
\begin{align}
	X^n_t =& X^n_0 + \int_{t_0}^t\frac{X^{n-2}_s}{g(s)} \left [ F_n X^2_s + A_n X_s + B_n(s)\right ] \ud s \nonumber\\
  &+ n \int_{t_0}^t \frac{X^{n-1}_s}{\sqrt{g(s)}} \sqrt{cX^2_s+dX_s+ e(s)}~\ud W_s \ ,
\end{align}
whose expectation readily provides the linear ordinary differential equation satisfied by the $n$-th order moment $\mu_n(t)=\ang{X^n_t}$ for $n\geq 1$
\begin{equation}
  g(t)\frac{\ud}{\ud t} \mu_n(t) = F_n \mu_n(t) +A_n \mu_{n-1}(t) + B_n(t) \mu_{n-2}(t)\ ,
  \label{eq:moments_diffeq}
\end{equation}
with boundary condition $\mu_n(t_0)=\ang{X^n_0}$. 
The coefficients read $F_n = na+\frac{1}{2}n(n-1)c$, $A_n = nb+\frac{1}{2}n(n-1)d$, $B_n(t) = \frac{1}{2}n(n-1)e(t)$, and we assume $\mu_0(t)=1$. 
If $\ang{X^k_0}$ for $k=1,\ldots,n$ is a finite quantity, the smoothness of $1/g(t)$ and $e(t)$ ensures the existence of a unique solution $\mu_n(t)$ over an arbitrary interval 
$D'=[t_i,t_f]\subseteq D$ with $t_0\in D'$. In terms of the monotonously increasing function $\tau(t)~=~\int_{t_0}^t 1/g(s) \ud s$, 
the solution reads
\begin{align}
  \mu_n(t) = &~\ue^{F_n \tau(t)} \left [ \ang{X^n_0} + \int_{0}^{\tau(t)} \ue^{-F_n \tau_1} A_n \tilde\mu_{n-1}(\tau_1) \ud \tau_1\right.\nonumber\\
  &\left .+\int_0^{\tau(t)} \ue^{-F_n \tau_1} \tilde B_n(\tau_1) \tilde\mu_{n-2}(\tau_1) \ud \tau_1 \right] ,
  \label{eq:mom_iter_sol}
\end{align}
where $\tilde\mu_n(\tau)=\mu_n(t(\tau))$ and $\tilde B_n(\tau)=B_n(t(\tau))$.
The previous expression lends itself to an expansion over $\ang{X^{n-j}_0}$, for $j=0,\ldots,n$ 
by iteratively substituting the moments entering the r.h.s. with their closed-form solutions.
We detail how to proceed for the simpler case $e(t)=e\geq0$.
We define type $A$ and type $B$ ``knots'' of order $k$ whose contributions are
\begin{equation*}
  \mathbb{A}_k=A_k \int_0^{\tau} \ue^{a_k \tau'} \ud \tau' \quad\mathrm{and}\quad
  \mathbb{B}_k=B_k \int_0^{\tau} \ue^{b_k \tau'} \ud \tau'\ ,
\end{equation*}
with $a_k = -(F_k-F_{k-1})$ and $b_k = -(F_k-F_{k-2})$.
We now consider ordered sequences of knots obtained applying the following rules:
a)~fix the order of the moment $n\in\{1,\ldots,N\}$;
b)~fix $j\in \{1,\ldots,n\}$;
c)~choose the first knot between $\mathbb{A}_n$ or $\mathbb{B}_n$;
d)~move rightward adding a new knot. $\mathbb{A}_k$ can be followed by either $\mathbb{A}_{k-1}$ or by $\mathbb{B}_{k-1}$, 
while $\mathbb{B}_k$ can be followed by either $\mathbb{A}_{k-2}$ or $\mathbb{B}_{k-2}$;~
e)~if $N_A$ and $N_B$ are the number of type $A$ and type $B$ knots, respectively, stop when $N_A+2N_B=j$.
We associate to every ordered sequence generated going through the previous procedure an integral
made of $N_A+N_B$ nested integrals.  
For instance, for $j=1$ the only admissible sequence is $\mathbb{A}_n$. 
For $j=2$, beside $\mathbb{B}_n$, we have to consider the sequence $\mathbb{A}_n \mathbb{A}_{n-1}$ giving the contribution 
\begin{equation*}
        \mathbb{A}_n \mathbb{A}_{n-1}=
	A_n A_{n-1} \int_0^{\tau} \ue^{a_n \tau_1}\int_0^{\tau_1} \ue^{a_{n-1} \tau_2} \ud \tau_2 \ud \tau_1\ .
\end{equation*}
When $j=4$ the following strings have to be taken into account
\begin{align*}
	&\mathbb{A}_n\mathbb{A}_{n-1}\mathbb{A}_{n-2}\mathbb{A}_{n-3},\quad
	\mathbb{A}_n\mathbb{A}_{n-1}\mathbb{B}_{n-2}                 ,\quad
	\mathbb{A}_n\mathbb{B}_{n-1}\mathbb{A}_{n-3}                 ,\quad\\
	&\mathbb{B}_n\mathbb{A}_{n-2}\mathbb{A}_{n-3}		     ,\quad
	\mathbb{B}_n\mathbb{B}_{n-2}.
\end{align*}
The special case $j=0$ is associated to a sequence with no knot whose contribution is equal to 1.
Once $n$ has been fixed, it is readily proved that every sequence is univocally determined retaining the label of the vertex while dropping the indexes.
For the case $j=4$ above strings reduce to $\mathbb{A}\mathbb{A}\mathbb{A}\mathbb{A},\mathbb{A}\mathbb{A}\mathbb{B},\mathbb{A}\mathbb{B}\mathbb{A},
\mathbb{B}\mathbb{A}\mathbb{A},\mathbb{B}\mathbb{B}$. We call $\Pi_{N_A N_B}$ the set of permutations 
with no repetition of $N_A$ type $A$ elements and $N_B$ type $B$ elements and $\pi_{N_A N_B}$ its generic element;
the compact notation $\Delta_n(\pi_{N_A N_B},\tau(t))$ identifies the $N_A+N_B$-dimensional integral contributing to the $n$-th moment and corresponding 
to the sequence of knots sorted according to $\pi_{N_A N_B}$.
In terms of the above quantities, the expression of $\mu_n(t)$ can be usefully rewritten in the compact form 
\begin{equation}
	\ue^{F_n \tau(t)} \sum_{j=0}^n \ang{X^{n-j}_0}\!\!\! \sum_{N_A+2N_B=j} \sum_{\Pi_{N_A N_B}}\!\!\!\!\Delta_n (\pi_{N_A N_B},\tau(t))~.
  \label{eq:general_expansion}
\end{equation}
A careful analysis of the quantity $\Delta_n(\pi_{N_A N_B},\tau(t))$ shows
that it can always be computed analytically in an algorithmic way~\footnote{We will
detail the derivation of Eqs.~\eqref{eq:general_expansion}
-\eqref{eq:expansion_algebraic}, as well as the algorithm allowing their numerical evaluation, in a forthcoming paper.}, which
makes the expansion~\eqref{eq:general_expansion} a powerful tool to exactly
compute $\mu_n$ up to an arbitrary order. 
Supposing that all the $a_k$ and $b_k$ involved in the expression of $\mu_n$ are non
vanishing \footnote{We do not consider in this work the case for an integer or semi-integer ratio between $\lvert a\rvert$ and $c$. 
This case can be dealt with by means of nested integrations by parts.}, 
the expansion~\eqref{eq:general_expansion} can be rewritten as
\begin{equation}
  \mu_n(t) = \sum_{j=0}^n c^n_j \ue^{F_{n-j}\tau(t)}\ ,
  \label{eq:expansion_algebraic}
\end{equation}
the $c_j^n$ being real possibly vanishing functions of
the $A_k$'s, $B_k$'s, $F_k$'s and $\ang{X^k_0}$. 
Above equation provides evidence of the typical scaling of the moments over time.
The multiple time scales emerging from the multiplicative noise process can be affected by varying the functional form of $g(t)$.
For a constant $g(t)=1$ \cite{Biro_Jakovac:2005}, we have $\tau=t-t_0$ and the $n$-th order moment is characterized by 
the superposition of $n$ exponentials with time constants $\{1/\lvert F_n\rvert,\ldots,1/\lvert F_1\rvert\}$.
When $g(t)=t$, as in \cite{Bottcher_etal:2006}, we have terms of the form
\begin{equation*}
	\ue^{F_{n-j}\tau(t)} =  t^{F_{n-j}}~t_0^{-F_{n-j}},
\end{equation*}
producing a power law time scaling of the moments.
More generally, for $g(t)=t^{\beta}$
($\beta\neq 1$) the time dependence turns out to be a stretched exponential with
stretching exponent $1-\beta$:
\begin{equation*}
  \ue^{F_{n-j}\tau(t)} = \ue^{F_{n-j}\frac{1}{1-\beta}\left( t^{1-\beta}-t_0^{1-\beta} \right)}~.
\end{equation*}
Eq.~\eqref{eq:expansion_algebraic} also allows us to gain insight into the nature
of the stochastic process described by the model~\eqref{eq:model}, both at finite $t$ and for $t$ approaching $t_\mathrm{lim}$.
As far as the PDF $p(x,t)$ associated to above process is concerned, in general we are not allowed
to draw any conclusion about its shape. However, an important exception is the case when, for $t\rightarrow t_\mathrm{lim}$, we have
a diverging $\tau$. Indeed, in terms of $\tau$ the PDF satisfies the FP equation
\begin{equation}
  \frac{\partial}{\partial \tau} \tilde p(x,\tau) \!=\! -\frac{\partial}{\partial x}
  \left[ D_1(x) \tilde p(x,\tau) \right] + \frac{1}{2} \frac{\partial^2}{\partial x^2}
  \left[ D_2(x) \tilde p(x,\tau) \right]\ ,
  \label{eq:FP}
\end{equation}
with $D_1(x) = a x + b$ and $D_2(x) = c x^2 + d x + e$ and initial condition $\tilde p(x,0)=\tilde{p}_0(x)$, for which  
the stationary solution $\tilde p_\mathrm{st}(x)$ can be computed analytically following \cite{Risken:1996,Biro_Jakovac:2005}. 
This solution provides evidence of the possible emergence of power law tails and Eq.~\eqref{eq:expansion_algebraic} precisely characterizes the way the moments 
converge to the stationary level.  
Indeed, the smoothness of $\tau$ as a function of $t$ implies $\lim_{t\rightarrow t_\mathrm{lim}^-}p(x,t)=\lim_{\tau\rightarrow +\infty}\tilde p(x,\tau)=\tilde p_\mathrm{st}(x)$.
For example, if $g(t)=(t-t_\mathrm{lim})^\beta$ with $\beta>1$,  the behaviour of $\tilde p_\mathrm{st}(x)$ 
emerging from the analysis of Eq.~(\ref{eq:FP}) applies to $p(x,t)$ when $t$ approaches $t_\mathrm{lim}$, 
while the moments scale according to
\begin{equation*}
	\ue^{F_{n-j} \tau(t)} = \ue^{F_{n-j} \frac{1}{1-\beta}\left[(t_0-t_\mathrm{lim})^{1-\beta}-(t-t_\mathrm{lim})^{1-\beta}\right]}\ . 
\end{equation*}
\begin{figure}[t]
  \begin{center}
    \includegraphics[width=.41\textwidth]{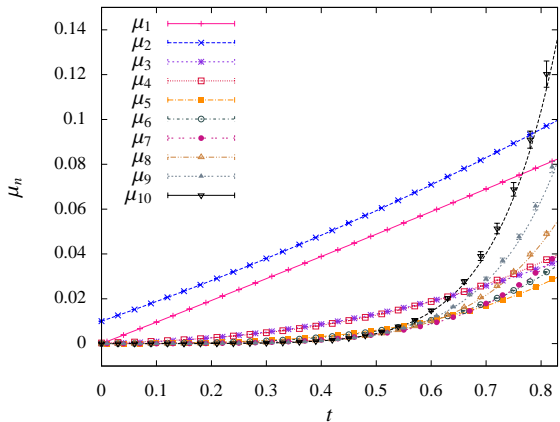}
  \end{center}
  \caption{(Color online) Scaling of the moments for $a=b=9.5 \times 10^{-2}$ and
  $c=d=e=8.3\times 10^{-2}$; $\tilde{p}_0(x)$ is a zero mean Gaussian with $\ang{X^2_0}=0.01$.
  The Monte Carlo estimates of the $\mu_n$'s within $68\%$ error bars are superimposed to the analytical curves, exhibiting full agreement.}
  \label{fig:PL1}
\end{figure}
We now discuss how our results can be employed to characterize the stochastic process described by the SDE \eqref{eq:model} for different choices of $a,b,c,d$, and $e$. 
Our analysis is essentially based on the sign of the factors $F_{n-j}$ appearing in \eqref{eq:expansion_algebraic}.
Indeed, $F_n$ is a convex function of $n$, depending only on $a$ and $c$, whose zeros are $n_0=0$ and $n_1=1-2a/c\in\mathbb{R}$. 
If $n_1<0$ all the moments diverge when  $t\rightarrow t_\mathrm{lim}^-$, while if $n_1>0$, all $\mu_n$'s for $n<n_1$ are convergent, otherwise not. 
For a convergent $\mu_n$, the estimate of the rate of convergence is provided by $\tau_\mathrm{max}=\max(1/\lvert a\rvert,1/\lvert F_n\rvert)$
which corresponds to the largest relaxation time in \eqref{eq:expansion_algebraic}.  
The cases $n_1=0,1,2$ have to be considered carefully, since they correspond to $a=c/2,0$ and $-c/2$, respectively, 
for which Eq.~(\ref{eq:expansion_algebraic}) does not apply (see \footnotemark[\value{footnote}]).\\
\emph{Case $a>0$ and $c>0$.} All the moments diverge, since $F_n>0$ $\forall n>0$. From \cite{Biro_Jakovac:2005} 
the stationary solution can be defined only if $0<a<c/2$ and $e>0$ and it is a generalized Student-$t$ with tail exponent $\nu=1+n_1$. 
For $a\geq c/2$ and finite $t$ all the moments are well-defined but no conclusions can be drawn about the exact form of the PDF. An example
of the latter case is shown in Figure~\ref{fig:PL1}, with all the $\mu_n$ diverging for large $t$.\\
\emph{Case $a=0$ and $e>0$.} If $c=0$, then also $d$ is 0 and Eq.~(\ref{eq:model}) describes an Arithmetic Brownian motion with time dependent coefficients.
If $c>0$, then $a_1=F_1=0$ and $\mathbb{A}_1=b\tau$, but integration by parts reveals that no moment converge, while the stationary solution is 
a power law with tail exponent $\nu=2$.\\ 
\emph{Case $a<0$, $c>0$, and $e>0$.} $F_n>0$ for $n>n_1$, thus only the first $n<n_1$ moments converge to a stationary level.
The special case $a=-c/2$ implies $n_1=2$, $b_2=F_2=0$ and $\mathbb{B}_2=e\tau$ and previous conclusions are unchanged.
Coherently the solution of the FP equation predicts the emergence
of power law tails with $\nu=1+n_1$. In Figure~\ref{fig:PL10_stretched} a case corresponding to $n_1 \simeq 9.9$ is shown 
for a stretched exponential time scaling, while Figure~\ref{fig:PL5}
corresponds to the case $g(t)=(t-t_\mathrm{lim})^2$.\\
\emph{Case $a\neq 0$, $c=0$, and $e>0$.} Eq.~(\ref{eq:model}) describes an Ornstein-Uhlenbeck process. $F_n$ becomes a linear function of $n$ and
the moments reach a stationary value only if $a<0$. For $a>0$ the Gaussian PDF has
time dependent unbounded mean and variance.\\
\emph{Case $e=0$ and $c>0$.} 
As above the boundedness of the moments can be deduced from the value of $n_1$ and for $a<0$ the stationary solution is an Inverted Gamma
with shape parameter $n_1>0$ and scale parameter $2\lvert b\rvert/c>0$. 
If $b>0$ the Inverted Gamma is defined for $x\in[0,+\infty)$, while for $b<0$ the support is $(-\infty,0]$. 
A similar situation occurs for $d-4ce=0$ and $d>0$, $c>0$, and $e>0$, but the support boundary point corresponds to $-d/(2c)$.

We can also deal with the more general case of a time dependent $e(t)$, for which an analysis of the $\tilde p_\mathrm{st}(x)$ 
can not be performed straightforwardly. Here we outline how to proceed for two cases discussed in the literature 
\cite{Friedrich_etal:2000,BorlandPRL:2002}, 
emerging in the context of financial time series analysis. 
Eq.~(6) in \cite{Friedrich_etal:2000} leads to a SDE belonging to the class~(\ref{eq:model}) for $g(t)=1$, implying $\tau=t-t_0$, $a=-4.4\times 10^{-1}$, $b=0$, 
$c=3.8\times 10^{-2}$, $d=3.04\times 10^{-3}$, and $e(t)=e+e'\ue^{\epsilon t}$, with $e=6.08\times 10^{-5}$, $e'=3\times 10^{-3}$, and
$\epsilon=-0.5$. The type B knot now splits into the sum of two contributions 
\begin{equation*}
	\mathbb{B}_k=B_k \int_0^{\tau} \ue^{b_k \tau'} \ud \tau' \quad\mathrm{and}\quad \mathbb{B}_k'=B_k' \int_0^{\tau} \ue^{b_k' \tau'} \ud \tau'\ ,
\end{equation*}
where $B_k'=\frac{1}{2}k(k-1)e'\ue^{\epsilon t_0}$ and $b_k'=b_k+\epsilon$.
The last two sums in~(\ref{eq:general_expansion}) have to be coherently modified as 
\begin{equation*}
	\sum_{N_A+2(N_B+N_{B'})=j} \sum_{~\Pi_{N_A N_B N_{B'}}}\Delta_n (\pi_{N_A N_B N_{B'}},t-t_0)\ ,
\end{equation*}
while Eq.~(\ref{eq:expansion_algebraic}) becomes
\begin{equation*}
  \mu_n(t) = \sum_{j=0}^n c^n_j(\epsilon,t_0) \ue^{(F_{n-j}+d_j^n \epsilon)(t-t_0)}\ ,
\end{equation*}
with $d_j^n\in\mathbb{N}$.

The second case we want to discuss is the model assumed in~\cite{BorlandPRL:2002} to describe the financial returns dynamics under the objective probability 
measure. It is readily proved that, defining $X_t=\ln{S_t/S_0}-\mu t$, Eq.~(2) in~\cite{BorlandPRL:2002} corresponds
to Eq.~(\ref{eq:model}) for $g(t)=t$, $a=b=d=0$, $c=(q-1)/[(2-q)(3-q)]$, $e(t)=et^{2/(3-q)}$ 
with $e=\sigma^2[\mathrm{c}(2-q)(3-q)]^{(q-1)/(3-q))}$, and $\ang{X^{n}_0}=0, \forall n\geq 1$;
$\sigma^2$ and $\mathrm{c}$ are positive constants, while the Tsallis entropic index $q$ belongs to $(1,5/3)$
to ensure the existence of mean and variance~\cite{Tsallis:1988}. The starting time $t_0=0$ is a singular point for $1/g(t)$ and we reproduce the correct
results in the limit $t_0\rightarrow 0^+$. Since $\ang{X^{n}_0}=0$ the only $j$ contributing to the sum in~(\ref{eq:general_expansion}) is $j=n$, while
$b=d=0$ implies that every type A knot is identically zero and so are all the odd moments. 
Thus, expansion~(\ref{eq:general_expansion}) reduces to 
$\ue^{F_n \tau(t)} \Delta_n(\pi_{0 \frac{n}{2}},\tau(t))$ with $\tau=\ln t - \ln t_0$ and $n=2p$, $p>1$. 
Now the type B knot reads
\begin{equation*}
	\mathbb{B}_k=\frac{1}{2}k(k-1)e~\ue^{\frac{2}{3-q}\ln t_0}\int_0^\tau \ue^{(b_k+\frac{2}{3-q})\tau'}\ud\tau'\ .
\end{equation*}
The same analysis leading to \eqref{eq:expansion_algebraic} allows us to conclude that
\begin{equation*}
	\mu_{2p}(t)=\sum_{j=0}^{p}c_j^{2p}t^{F_{2(p-j)}+\frac{2j}{3-q}}t_0^{-\left[F_{2(p-j)}+\frac{2(j-p)}{3-q}\right]}\ , 
\end{equation*}
where the $c_j^{2p}$ are all non-vanishing constants. The expression in the square brackets governs the limiting behaviour of $\mu_{2p}$ when $t_0\rightarrow 0^+$.
As a function of $j$, $F_{2(p-j)}+2(j-p)/(3-q)$ is convex and for $j=p$ its value is zero, so that we only have to check the behaviour for $j=0$.
Indeed, if $F_{2p}+2p/(q-3)>0$ then $\lim_{t_0\rightarrow 0^+}\mu_{2p}=+\infty$, otherwise all the exponents of $t_0$ are non-negative and no divergence is possible.
But recalling the expression of $c$, we find $q>(2p+3)/(2p+1)$,
recovering the condition required in~\cite{BorlandPRL:2002} to obtain a divergent $2p$-th order moment. It is worth noticing that the previous 
conclusions can be readily rephrased in terms of the Hurst exponent $H=1/(3-q)$ as it has been done in \cite{McCauley_etal:2007}.

In conclusion, the processes described by Eq.~\eqref{eq:model} emerge in the study of many complex systems.
An explicit expression for the associated PDF has not, however, been available for finite times.
In this Report we provided a characterization in terms of the moments, deriving closed-form expressions at all orders
and for all times.
These results, revealing a simple relationship between the
time dependence of Eq.~\eqref{eq:model} and the scaling behaviour of the moments provides a better understanding of how these processes evolve
with time, possibly but not necessarily toward a stationary state and provide a closed-form relation between the model parameters and
their relaxation times. We also believe that these results can improve the statistical analysis of historical time series. 
Indeed, the analytical expressions allow to directly fit the time scaling of the empirical moments,
providing a clear and simple way to fix the model parameters. 
Moreover, the knowledge of the moments is crucial to exploit analytical approximation to the full PDF associated to Eq.~\eqref{eq:model},
such as the Edgeworth or more general expansions~\cite{Gaztanaga_etal:2000}, even though recovering the PDF is a problem whose solution
is in general not unique.

\end{document}